\documentclass[aps,prl,twocolumn,superscriptaddress,showpacs,amsmath,amssymb]{revtex4}
\usepackage{graphicx}
\usepackage{bm}
\usepackage{amssymb}
\newcommand{\be}{\begin{equation}}
\newcommand{\ee}{\end{equation}}
\newcommand{\bea}{\begin{eqnarray}}
\newcommand{\eea}{\end{eqnarray}}

\newcommand{\br}{{\bf r }}

\newcommand{\bk}{{\bf k}}

\def\lsim{\mathrel{\rlap{\lower4pt\hbox{\hskip1pt$\sim$}}
    \raise1pt\hbox{$<$}}}         

\def\gsim{\mathrel{\rlap{\lower4pt\hbox{\hskip1pt$\sim$}}
    \raise1pt\hbox{$>$}}}         

\begin{document}

\textheight 24 cm

\title{Creation of two vortex-entangled beams in a vortex beam collision with a plane wave}
\author{Igor P. Ivanov}
\affiliation{IFPA, Universit\'{e} de Li\`{e}ge, All\'{e}e du 6 Ao\^{u}t 17, b\^{a}timent B5a, 4000 Li\`{e}ge, Belgium}
\affiliation{Sobolev Institute of Mathematics, Koptyug avenue 4, 630090, Novosibirsk, Russia}

\begin{abstract}
Physics of photons and electrons carrying orbital angular momentum (OAM) is an exciting field of research in quantum optics
and electron microscopy.
Usually, one considers propagation of these vortex beams in a medium or external fields and their absorption or scattering on fixed targets.
Here we consider instead a beam-beam collision. We show that elastic scattering of a Bessel vortex beam 
with a counterpropagating plane wave naturally leads to two vortex-entangled outgoing beams. 
The vortex entanglement implies that the two final particles are entangled not only in their orbital helicities but also in 
opening angles of their momentum cones.
Our results are driven by kinematics of vortex-beam scattering and apply to particle pairs of any nature:
$e\gamma$, $e^+e^-$, $ep$, etc.
This collisional vortex entanglement can be used to create pairs of OAM-entangled particles of different nature, 
and to transfer a phase vortex, for example, from low-energy electrons to high-energy protons. 
\end{abstract}


\maketitle

\textit{Introduction.---} Laser beams carrying non-zero orbital angular momentum (OAM) are well-known and routinely used in optics, \cite{OAM,OAMreview}.
Wavefronts of such a beam are not planes but helices, and each photon in this light field (a twisted photon) carries 
a well-defined OAM quantized in units of $\hbar$.
Applications of twisted photons range from microscopy to astrophysics, \cite{TP}. 
They are also interesting for quantum information science because of the high dimensionality of the OAM state space
and the possibility to create OAM-entangled pairs of photons, \cite{zeilinger,high-dimensional}.

Wave-fronts with phase vortices can exist for electrons and other particles as well.
Recently, following the suggestion of \cite{bliokh2007}, electron beams carrying OAM were experimentally demonstrated,
first using phase plates \cite{twisted-electron} and then with fork diffraction holograms \cite{twisted-electron2}.
Such electrons carried kinetic energy as high as 300 keV and the orbital quantum number up to $m\sim 100$.
Exact solution of the Dirac equation representing a relativistic electron vortex was recently derived in \cite{bliokh2011}.

Experimental realization of vortex beams opens up the possibility to study head-on beam-beam scattering,
with one or both colliding beams carrying OAM. Using sufficiently energetic twisted electrons and photons in $e\gamma$, $ee$, $e^+e^-$,
$ep$ or electron-nucleus collisions, one can probe quantum electrodynamics and perhaps hadronic processes
in a novel way,  \cite{ivanov2011}.
In addition, the Compton backscattering of optical twisted photons 
from an ultrarelativistic electron beam was suggested in \cite{serbo} to generate twisted photons 
in the GeV energy range.

All such processes share certain universal features, which are driven by kinematics of free-space beam-beam collision
and are not sensitive to its microscopic nature. Some of them were studied in \cite{ivanov2011,IS2011}.
One particular finding was the necessity of the use of {\em orbital helicity}, the OAM projection on the average propagation 
direction (which is different for each particle) rather than on a fixed common axis.
This notion becomes especially important for high-energy and for high-angle (strongly non-collinear) scattering,
as it removes strong ``instrumental OAM'' effects caused by an unfortunate choice 
of the reference axis and leaves us with the ``intrinsic OAM'' of the twisted state.
The usage of orbital helicity (without calling it so) was also advocated in \cite{Osorio2009}.
In the present paper whenever we refer to OAM we actually mean the orbital helicity.

In this Letter we explore another universal feature of vortex beam collisions.
We study whether the outgoing wavefronts contain vortices and what are the corresponding orbital helicities $m_1$ and $m_2$. 
This issue has never been addressed in a generic set-up; in previous calculations of vortex beam scattering at least one of the final
particles was assumed to be a plane wave.
We find that the final particles are {\em vortex-entangled}: that is, they are entangled both in the $(m_1,m_2)$-space 
and in the space of opening angles.
Since for non-collinear scattering there is no total orbital helicity conservation law, 
selecting one specific $m_1$ does not lead to a unique $m_2$.
However by using the other degree of freedom, the opening angle, one can collapse the second particle to a more-or-less definite $m_2$,
with important practical implications.

Note that similar issues arise in the context of OAM-entangled photon pairs production by spontaneous parametric down-conversion (SPDC) in a non-linear medium, 
see for example \cite{Osorio2009,discussion}. In the particle language, such medium induces decay of the pump photon into two photons of lower energy. 
Although a convenient source of OAM-entangled photon
pairs, this process is specific to photons and its dynamics is defined (and limited) by the exact properties of the crystal slab, including its thickness. 
In contrast to that, the entanglement discussed in this Letter is universal (not specific to particle species), it spontaneously occurs 
in a free-space collision (and not induced by a medium), and it allows one to study its behavior in the entire available kinematical range.

\textit{Describing twisted states.---} 
To simplify the calculations, we describe the twisted states as paraxial Bessel beams.
We consider the scalar case only (polarization degrees of freedom can be straightforwardly incorporated
in the paraxial approximation) 
and use the conventions of \cite{serbo}.
All transverse vectors are given in bold, while three-vectors will be indicated by the arrow.
A Bessel twisted state is a non-plane wave solution of the free wave equation
with a definite frequency $\omega$,
longitudinal momentum $k_z$, modulus of the transverse momentum
$|\bk|=\varkappa$ and a definite $z$-projection of orbital angular momentum $m$.
When written in cylindric coordinates $r, \varphi_r, z$, it has form 
$|\varkappa, m\rangle = e^{-i\omega t + i k_z z} 
\psi_{\varkappa m}(\br)$, with
$\psi_{\varkappa m}(\br) = e^{i m \varphi_r} J_{m}(\varkappa r) \sqrt{\varkappa/2\pi}$,
where $J_m(x)$ is the Bessel function. 
A twisted state can be represented as a superposition of plane
waves
 \be
|\varkappa,m\rangle = e^{-i\omega t + i k_z z} \int {d^2\bk
\over(2\pi)^2}a_{\varkappa m}(\bk) e^{i\bk \br}\,,
 \label{twisted-def}
  \ee
where
 \be
a_{\varkappa m}(\bk)= (-i)^m
e^{im\varphi_k}\sqrt{2\pi}\;{\delta(|\bk|-\varkappa)\over
\sqrt{\varkappa}}\,.\label{a}
  \ee
Thus, the allowed momenta lie on the edge of a cone with the opening angle $\arctan(\varkappa/k_z)$.
More properties of twisted states, including their
normalization and phase space density, can be found in
\cite{serbo,ivanov2011}. Here we only note that Eq.~(\ref{twisted-def}) in fact describes the passage
from plane waves to twisted particles in description of a scattering process.

A Bessel state $|\varkappa,m\rangle$ with fixed $\varkappa$ is non-normalizable in the transverse plane.
A much better approximation to physically realizable states such as Bessel-Gaussian or aperture-limited beams
is given by a fixed-$m$ superposition of Bessel states 
\be
|\varkappa_0,\sigma;m\rangle = \int d\varkappa \, f(\varkappa) |\varkappa,m\rangle\,,\label{WP}
\ee
with a properly normalized weight function $f(\varkappa)$ peaked at $\varkappa_0$ and 
having width $\sigma$. This state is normalizable (and localized) in the transverse plane
and is assumed to be monochromatic ($k_z$ is supposed to vary with $\varkappa$ so that the energy is constant).
Properties of such states and their important role in resolving
the non-forward-to-forward paradox in Bessel beam scattering was discussed in \cite{IS2011}.


\textit{Vortex beam scattering.---} 
To describe scattering of the Bessel state $|\kappa,m\rangle$ and a plane-wave,
we start with a generic elastic two-particle scattering
in the usual plane wave basis.
The initial particle four-momenta are denoted as $k$ and $p$, the final four-momenta are $k_1$ and $k_2$.
The scattering matrix element of this process is represented as
\be
S_{PW} = i(2\pi)^4\delta^{(4)}(k + p - k_1 -k_2) {\cal M}\,, \label{SPW}
\ee
where the amplitude ${\cal M}$ is calculated according to the standard Feynman rules.
Then we pass from the plane-wave to twisted scattering by
convoluting (\ref{SPW}) with the weight function $a_{\varkappa_i m_i}$, (\ref{a}), 
for each initial Bessel-beam twisted state, 
and with $a^*_{\varkappa_f m_f}$ for each final twisted state.
For example, if only one initial particle is twisted, the scattering matrix element
is $S_{tw} = \int d^2\bk a_{\varkappa m}(\bk) S_{PW} / (2\pi)^2$.
Owing to the delta-function inside $a_{\varkappa m}$, this representation
contains only one integration with respect to the azimuthal angle $\varphi$ of the transverse momentum $\bk$.
This integration is eliminated by the transverse momentum delta-function in $S_{PW}$, (\ref{SPW}).
If we assume for simplicity that $\vec p = (0,0,p_z)$ and introduce the well-defined transverse vector
$\bk_{12} \equiv \bk_1 + \bk_2$ with modulus $k_{12}$ and azimuthal angle $\varphi_{12}$, then
\bea
S_{tw}& \propto&\int {d^2 \bk \over (2\pi)^2} a_{\varkappa m}(\bk) \delta^{(2)}(\bk-\bk_1-\bk_2) {\cal M}(\bk)\nonumber\\
&& = {(-i)^m \over(2\pi)^{3/2}} e^{im\varphi_{12}}{ \delta(\varkappa-k_{12}) \over \sqrt{\varkappa}} {\cal M}(\bk_{12})\,.\label{Stw1}
\eea
Note that the scattering amplitude ${\cal M}$ itself is not integrated but is just taken at a specific value of the initial momentum.

Do the wavefronts of the two outgoing particles contain phase vortices?
Eq.~(\ref{Stw1}) provides the answer to this question.
Let us first suppose that the final momentum $\bk_2$ with modulus $|\bk_2|$ and azimuthal angle $\varphi_2$ 
is fixed (the second particle is projected on a plane wave). 
Then, the final momentum of the first particle $\bk_1$
is not uniquely defined but belongs to a circle of radius $\varkappa$ around the point $-\bk_2$.
Its modulus changes in the interval $\bigl||\bk_2|-\varkappa\bigr| \le |\bk_1| \le |\bk_2|+\varkappa$,
and for a given $|\bk_1|$ its azimuthal angle $\varphi_1$ takes two values:
$\varphi_1 = \varphi_2 \pm \arccos\left({\varkappa^2 - \bk_1^2 - \bk_2^2 \over 2 |\bk_1| |\bk_2| } \right)$.
The azimuthal angle $\varphi_{12}$, which parametrizes the points on the circle, also takes two values
$\varphi_{12} = \varphi_2 \pm \arccos\left({\varkappa^2 + \bk_2^2 - \bk_1^2 \over 2 \varkappa |\bk_2| } \right)$,
and the plus-minus signs in these two expressions are correlated.
When $|\bk_1|$ spans the allowed interval, the angle $\varphi_{12}$ covers the entire circle,
although the intensity of the scattering is modulated by the amplitude ${\cal M}$.

Note that the scattering amplitude for $\bk_1 = -\bk_2$ is exactly zero
for any value of $\varkappa$. Passing to the three-dimensional vectors,
one can define the axis $\vec n_1 \| \langle \vec k \rangle + \vec p - \vec k_2$,
where $\langle \vec k \rangle \equiv \langle \varkappa,m|\vec k| \varkappa,m\rangle$,
and claim that scattering exactly in this direction in absent
even of the initial twisted state is represented by a transversely localized state (\ref{WP}).
Therefore, $\vec n_1$ represents 
the {\em direction of the phase vortex} (the line of zero intensity and undefined phase).
It is, therefore, natural to expect that at fixed $\vec k_2$ the final state of the first particle 
can be approximated by the twisted state $|\varkappa_1,m_1\rangle$ 
defined with respect to this direction, with $\varkappa_1 \approx \varkappa$ 
and with orbital helicity $m_1 \approx m$.
For the specific kinematics of the Compton backscattering, 
this expectation was confirmed in \cite{IS2011}. 

It is clear that if one fixed the final momentum $\vec k_1$ of the first particle, one would rederive similar conclusions for $\vec k_2$:
the outgoing wave of the second final particle would contain a phase vortex in the direction of 
$\vec n_2 \| \langle \vec k \rangle + \vec p - \vec k_1$.
Therefore, there can be no unambiguous assignment which of the final particles is twisted and which is not.

\begin{figure}[!htb]
   \centering
\includegraphics[width=7cm]{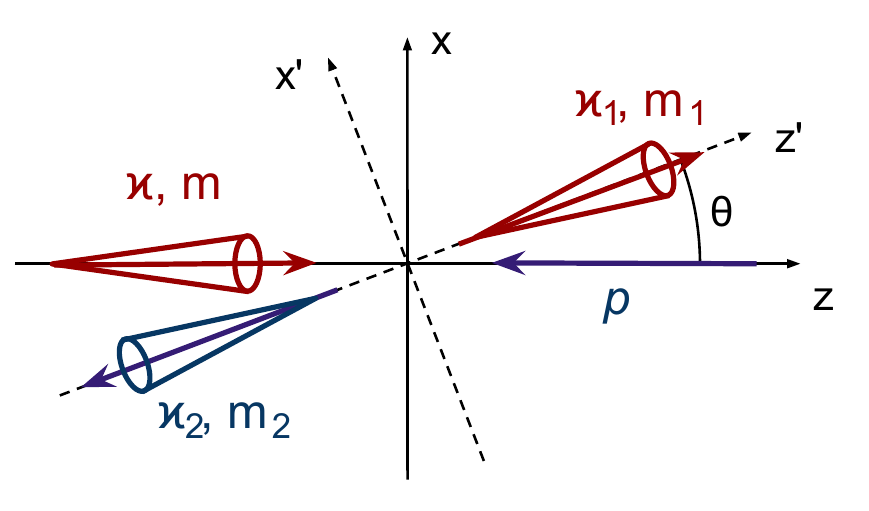}
\caption{(Color online.) Kinematics of the triple-twisted scattering in the c.m.s.}
   \label{fig1}
\end{figure}

\textit{Triple-twisted scattering.--- }The ``plane wave/twisted state'' entanglement derived above 
rests on the assumption that one of the final particles is a plane wave.
Below we show that, in general, 
the final state in this elastic collision can be represented by an {\em entangled state of two vortex beams}.
The final twisted particles are entangled not only through $m$'s, but also through $\varkappa$'s,
and the previously found ``plane wave/twisted state'' entanglement is just a particular case of this general result.

Let us again consider the elastic collision of a Bessel state $|\varkappa,m\rangle$ with the plane wave 
with momentum $\vec p$ and represent the two final particles as Bessel states $|\varkappa_1,m_1\rangle$ and $|\varkappa_2,m_2\rangle$.
Kinematical conventions are shown in Fig.~\ref{fig1}.
In order to simplify the calculations, we consider this process in the c.m.s. frame, where $\langle\vec k\rangle + \vec p = 0$, 
and choose the common quantization axis $z'$, characterized by the polar angle $\theta$, for both final particles.
The two axes, $z$ and $z'$, define the scattering plane; working
in this plane, we choose axis $x$ as the one orthogonal to $z$, and axis $x'$ as orthogonal to $z'$.
Without loss of generality, we assume that the azimuthal angle $\varphi$ for the initial twisted state is measured
from axis $x$, while the azimuthal angles $\varphi_1$ and $\varphi_2$ of the final twisted states (which lie in the
plane orthogonal to $z'$) are measured from the axis $x'$.
The average values of the final momenta $\langle\vec k_1\rangle$ and $\langle\vec k_2\rangle$
are parallel to $z'$; we denote their difference as $q \equiv |\langle\vec k_1\rangle| - |\langle\vec k_2\rangle| = k_{1z'}+k_{2z'}$.

The scattering matrix element for such a triple-twisted scattering, $S_{3tw}$, is equal to
\be
\int\!\! {d^2 \bk \over (2\pi)^2} {d^2 \bk_1 \over (2\pi)^2} {d^2 \bk_2\over (2\pi)^2}  
a_{\varkappa m}\!(\bk) a^*_{\varkappa_1 m_1}\!(\bk_1)   a^*_{\varkappa_2 m_2}(\bk_2)
S_{PW}\,.\label{S3tw}
\ee
The three-dimensional delta-function $\delta^{(3)}(\vec k + \vec p - \vec k_1 - \vec k_2)$ inside $S_{PW}$ eliminates
all three integrals over the azimuthal angles $\varphi$, $\varphi_1$ and $\varphi_2$ setting them to certain values.  
The amplitude ${\cal M}$ does not affect the integral and is just taken at the momenta $\vec k$, $\vec k_1$ and $\vec k_2$
corresponding to these azimuthal angles. In order to show the key features of the results, we now assume 
that the amplitude ${\cal M}$ is a slowly varying function, and in the paraxial approximation
we take ${\cal M}(\vec k_i) \approx {\cal M}(\langle\vec k_i\rangle) \equiv {\cal M}_0$ out of the integral.
Then, the integral can be computed exactly, and the triple-twisted scattering matrix element equals
\begin{widetext}
\bea
S_{3tw} &=& i{\delta(E_f-E_i) \over \sqrt{2\pi}}\cdot 
i^{m_1+m_2-m}{2\over \Delta}\sqrt{{\varkappa_1\varkappa_2 \over \varkappa}}
{\cos[m \varphi^* - (m_1-m_2)\tilde\varphi^*]\cdot
\cos[m_1\delta_1 + m_2\delta_2] \over \sqrt{\sin^2\theta - \sin^2\xi}}\cdot {\cal M}_0\,.\label{master4}
\eea
\end{widetext}
Here we introduced parameter $\xi$ via $\sin\xi = q/\varkappa$, $|\xi| < \theta$, and
the following angles
\be
\varphi^* = \arccos\left({\sin\xi \over \sin\theta}\right)\,,\quad 
\tilde\varphi^* = \arccos \left({\tan\xi  \over \tan\theta}\right)\,.
\ee
Besides, $\Delta$ is the area of the triangle with sides $\varkappa_1$, $\varkappa_2$, and $\tilde\varkappa \equiv \varkappa\cos\xi$,
and $\delta_1$ and $\delta_2$ are two of its inner angles:
$\cos\delta_{1,2} = (\tilde\varkappa^2 + \varkappa_{1,2}^2 - \varkappa_{2,1}^2)/2\tilde\varkappa \varkappa_{1,2}$.
The results of \cite{IS2011} for the case when one of the final particles is a plane wave can be recovered from (\ref{master4})
in the case of $m_2 = 0$ and $\varkappa_2 \to 0$, using the following relations, \cite{ivanov2011}:
$|PW(\bk_2)\rangle = \lim_{\varkappa_2 \to 0} \sqrt{2\pi/\varkappa_2} |\varkappa_2,0\rangle$ and  
$\lim_{\varkappa_2 \to 0} \Delta^{-1} = 4\pi \delta(\tilde\varkappa^2-\varkappa_1^2)$.

\begin{figure}[!htb]
   \centering
\includegraphics[width=6cm]{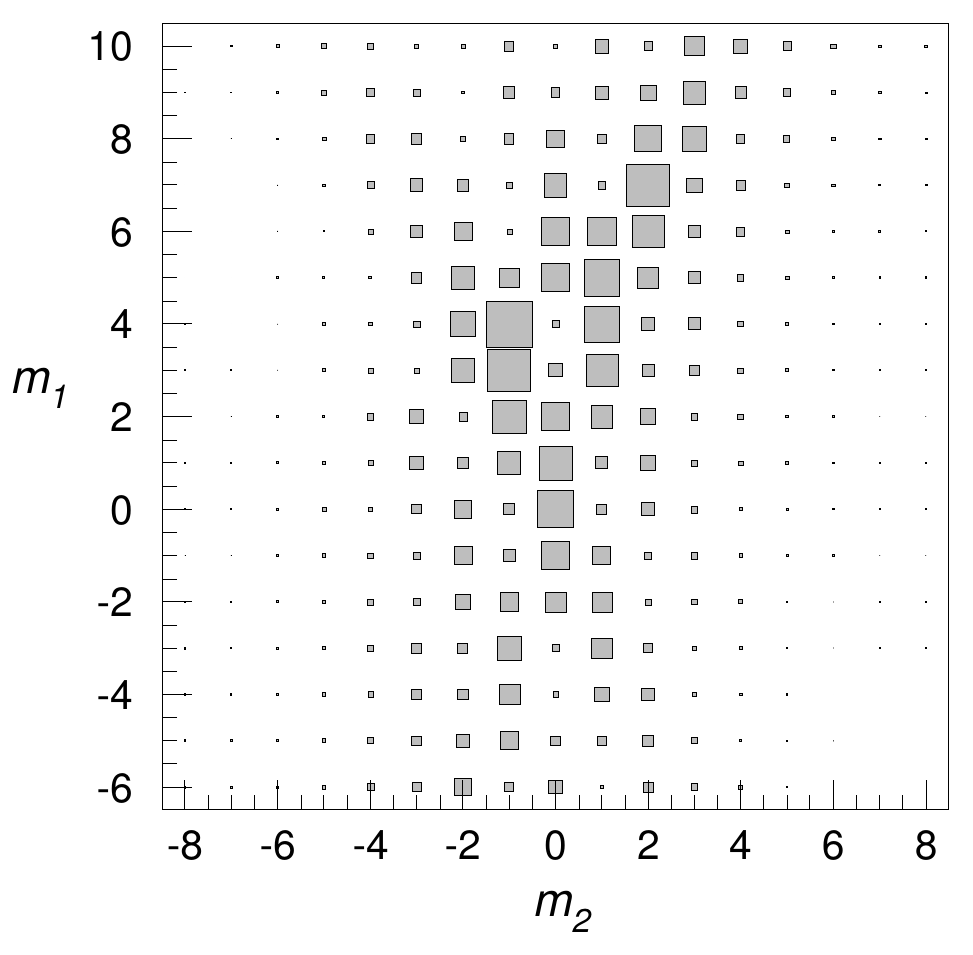}
\caption{Relative intensity of scattering on the $(m_1,m_2)$-plot integrated over $q$ 
for $\varkappa_0 = \varkappa_{01}= 2\varkappa_{02}$, $\sigma_i = \varkappa_i/5$, $m = 5$, and scattering angle $\theta = 0.2$.}
   \label{fig2}
\end{figure}

Our main result (\ref{master4}) allows us to characterize the entanglement of the two final particles.
They are entangled not only in the $(m_1,m_2)$ space,
but also in the $(\varkappa_1,\varkappa_2)$ space.
Existence of the triangle with sides $\varkappa_1$, $\varkappa_2$, and $\tilde\varkappa$ means that
the allowed values of $(\varkappa_1,\varkappa_2)$ lie inside the stripe defined by
\be
|\varkappa_1-\varkappa_2|\le \tilde\varkappa \le \varkappa_1+\varkappa_2\,.\label{stripe}
\ee
As for the $m_1$- and $m_2$-distributions, they are infinitely wide for pure Bessel states.
However if the initial and final twisted states are taken as transversely localized states (\ref{WP})
with peak positions $\varkappa_{0i}$ and widths $\sigma_i$,
then the orbital helicity distributions
become rather narrow with a typical width of ${\cal O}(\varkappa_{0i}/\sigma_i)$.
Their shapes also depend on the location of the central point $(\varkappa_{01},\varkappa_{02})$ in the 
stripe of allowed values (\ref{stripe}).
For illustration, we show in Fig.~\ref{fig2} a typical $(m_1,m_2)$-distribution of the scattering intensity as a proportional box plot
at $\theta = 0.2 \approx 11^\circ$
for $m=5$ and for an asymmetric choice of $\varkappa$'s:  $\varkappa_0 = \varkappa_{01}= 2\varkappa_{02}$ (the overall scale is arbitrary), 
$\sigma_i = \varkappa_{0i}/5$.
This scattering intensity is obtained by squaring the $f_i$-weighted $S$-matrix element (\ref{master4}) 
and integrating the result over the allowed region of $q$.
The $(m_1,m_2)$-correlations seen in Fig.~\ref{fig2} arise due to the orbital helicity entanglement of the two vortex beams.

The plot also illustrates the absence of the strict total orbital helicity conservation rule, although it possesses a sizable ``ridge'' near $m_1-m_2 = m$.
Another feature which is visible in the plot is that the final particle with the larger $\varkappa$ tends to carry larger orbital helicity.
This property can be read off (\ref{master4}). If $\varkappa_{2} \ll \varkappa_{1} \approx \varkappa$, 
then $\delta_{1} \ll 1$, while $\delta_2$ can be large. Therefore, $m_1$ can be large, while any non-zero $m_2$ will be suppressed by 
the second cosine after integration over certain $\varkappa_i$ domains. 
In the limit of vanishing $\varkappa_{2}$ or $\varkappa_{1}$ one recovers the ``plane wave/twisted state''
entanglement discussed above.

\textit{Discussion.---} 
Our results bear several implications.
First, we showed that elastic scattering of a vortex beam with a plane wave spontaneously leads to two vortex-entangled beams.
This can be used to create OAM-entangled pairs of particles of different nature (electro-photon, electron-proton, etc.), 
including those for which vortex beams are not yet available.
So far, only photon pairs entangled through OAM were created, \cite{zeilinger,high-dimensional}.
Experimental study of this entanglement is feasible with today technology, for example, by colliding electrons 
from two electron microscopes in the common focal spot and detecting the orbital helicity of scattered electrons.
Current discussion about OAM conservation during SPDC \cite{discussion} might also benefit from these experiments. 

Second, by colliding a vortex electron beam with a high-energy proton beam with sufficient transverse coherence
and by filtering only scattered electrons with small $\varkappa$, one can create energetic twisted protons.
This would be a unique possibility to generate high-energy particles carrying OAM, because fork diffraction grating 
or phase plates are of no use for high-energy (GeV range) particles.
Due to strong momentum imbalance, the kinematics of this process in the laboratory frame differs from the c.m.s. 
example considered here. However for small proton scattering angles the integral (\ref{S3tw}) can also be
calculated analytically, and the OAM-entanglement pattern can be studied.

Third, if a vortex electron beam experiences multiple scattering off stray atoms or other particles during its propagation, 
its twisted state can deteriorate. From the calculational point of view this process is different from 
propagation of twisted light through a turbulent medium or from motion of twisted electrons 
through chaotic electromagnetic fields. In contrast to scattering in an external potential, 
the scattering matrix element for a two plane wave collision contains the
three-dimensional delta-function describing momentum conservation, (\ref{SPW}).
Therefore, projecting the initial and final particles on twisted states will lead to different results in these two cases.

\textit{In summary}, we found that generic elastic scattering of a vortex beam with a plane wave leads to
two vortex-entangled outgoing beams. This feature was ignored in all previous calculations of vortex beam scattering,
\cite{ivanov2011,serbo,IS2011}.
The final particles are naturally entangled both through their orbital helicities $m_i$ 
and through $\varkappa_i$.
These results are driven by kinematics of vortex-beam scattering and apply to particles of any nature
($e\gamma$, $e^+e^-$, $ep$, etc).
They can be used to create pairs of OAM-entangled particles of different nature, to transfer a phase vortex
to high-energy particles (for example, protons), and are important for manipulating the electron vortex beams. 

\textit{Acknowledgements.---} 
I am grateful to V.~Serbo for stimulating discussions and suggestions.
This work was supported by the Belgian Fund F.R.S.-FNRS via the
contract of Charg\'e de recherches, and in part by grants of the Russian
Foundation for Basic Research 11-02-00242-a
and NSh-3810.2010.2.

\end{document}